\documentclass[11pt]{article}
\usepackage{amsmath,amssymb,color,epsf}
\usepackage{amssymb,slashed,latexsym,amsmath,multirow,color}
\usepackage{graphics}
\makeatletter
\@addtoreset{equation}{section}
\makeatother

\usepackage{amsfonts}

\newcommand{\hoch}[1]{$\, ^{#1}$}

\usepackage{float}

\usepackage{soul}

\usepackage{amsfonts}

\usepackage{tikz}
\usetikzlibrary{matrix,arrows,decorations.pathmorphing}
\usepackage{tikz-cd}

\usepackage{tikz}
\usetikzlibrary{shapes,shadows,arrows}
\usepackage[font=footnotesize,labelsep=newline,labelfont=sc,justification=centering,position=top]{caption}

\usepackage{tikz}
\usetikzlibrary{backgrounds}
\usetikzlibrary{shapes,shadows,arrows}
\tikzstyle{every picture}=[level distance = 8mm, baseline=-0.5ex]
\tikzstyle{prop}=[shape=circle,minimum size=6mm, draw=black!80, fill=green!30]

\usepackage{float}

\newcommand{\be}{\begin{equation}}
\newcommand{\ee}{\end{equation}}
\newcommand{\bea}{\setlength\arraycolsep{2pt} \begin{eqnarray}}
\newcommand{\eea}{\end{eqnarray}}
\newcommand{\nn}{\nonumber}

\usepackage{hyperref}

\newcommand{\ft}[2]{{\textstyle\frac{#1}{#2}}}
\def\rmi{{\rm i}}
\newsavebox{\uuunit}
\sbox{\uuunit}
    {\setlength{\unitlength}{0.825em}
     \begin{picture}(0.6,0.7)
        \thinlines
        \put(0,0){\line(1,0){0.5}}
        \put(0.15,0){\line(0,1){0.7}}
        \put(0.35,0){\line(0,1){0.8}}
       \multiput(0.3,0.8)(-0.04,-0.02){12}{\rule{0.5pt}{0.5pt}}
     \end {picture}}

\newcommand{\SU}{\mathop{\rm SU}}

\def\be{\begin{equation}}
\def\ee{\end{equation}}
\def\ba{\begin{array}}
\def\ea{\end{array}}
\def\bea{\begin{eqnarray}}
\def\eea{\end{eqnarray}}
\def\bd{\begin{displaymath}}
\def\ed{\end{displaymath}}

\def\nn{\nonumber}

\def\a{\alpha}

\def\g{\gamma}

\def\d{\delta}

\def\e{\epsilon}

\def\p{\psi}

\def\l{\lambda}
\def\L{\Lambda}
\def\m{\mu}
\def\n{\nu}
\def\r{\rho}
\def\s{\sigma}

\def\o{\omega}
\def\O{\Omega}

\def\nn{\nonumber}
\def\cD{\mathcal{D}}
\def\cN{\mathcal{N}}
\def\cA{\mathcal{A}}
\def\cL{\mathcal{L}}
\def\cF{\mathcal{F}}

\begin{document}

\begin{flushright}
\hfill{ \
MIFPA-14-06 }
\end{flushright}
\vskip 1.2cm

\begin{center}
{\Large \bf $R^n$ Extension of Starobinsky Model in Old Minimal Supergravity}
\end{center}
\vspace{25pt}
\begin{center}
{\Large {\bf }}

\vspace{10pt}
{Mehmet Ozkan$^1$ and Yi Pang$^2$}

\vspace{10pt}

\vskip .2truecm  \centerline{{\small $^1$1Centre for Theoretical Physics, University of Groningen,}}
      \centerline{{\small  Nijenborgh 4, 9747 AG Groningen, The Netherlands}}
\vskip .2truecm
{email: {\tt m.ozkan@rug.nl}}
\vskip .2truecm
\centerline{{\small  $^2$ George and Cynthia Woods Mitchell
Institute for Fundamental Physics and Astronomy,}} \centerline{\small Texas
A\&M University, College Station, TX 77843, USA}  \vspace{6pt}

{email: {\tt pangyi1@physics.tamu.edu}}
\vspace{40pt}
\vspace{10pt}

\hoch{} {\it }

\vspace{10pt}

\underline{ABSTRACT}
\end{center}
We provide a succinct way to construct the supersymmetric completion of $R^n$ $(n\ge3)$ in components using superconformal formulation of old minimal supergravity. As a consequence, we obtain the polynomial $f(R)$ supergravity extending the supersymmetric Starobinsky model to any higher power of $R$. The supersymmetric vacua in polynomial $f(R)$ supergravity are studied. We also present the $R^n$ extended Minimal Volkov-Akulov-Starobinsky Model.

\vspace{15pt}

\thispagestyle{empty}

\vspace{15pt}

 \vfill

\thispagestyle{empty}
\voffset=-40pt

\newpage

\tableofcontents


\section{Introduction}
If the perturbations during inflation \cite{Lyth:1998xn} are originated by the same field driving inflation, then the recent Planck data on the cosmic microwave background (CMB) radiation anisotropies have severely constrained the models of single-field inflation \cite{Ade:2013uln}. Successful models have to predict a significant red tilt in the power spectrum of the scalar curvature perturbation, measured by the spectral index $n_s=0.960\pm 0.007$, and a low enough amount of tensor perturbations quantified by the current bound on the tensor-to-scalar ratio, $r<0.08$. One of the models which better passes these constraints is the $R^2$ Starobinsky model \cite{Starobinsky:1980te,Mukhanov:1981xt}. It is described by the action ($M_{\rm pl}=1$)
\be
S=\frac12\int d^4x\sqrt{-g}\Big(R+R^2/(6M^2)\Big),
\ee
where the positivity of the coefficient in front of $R^2$ term is required to avoid instabilities. The inflaton with mass $M$ is given by the spin-0 part of the metric and thus has a clear geometric origin. CMB data determines the mass parameter $M$ to be $M=(3.0\times 10^{-6})\frac{50}{N_e}$, with $N_e$ being the e-folding number.

It is natural to consider embedding Starobinsky model in supergravity since supersymmetry is considered to be the leading proposal of new physics beyond the Standard Model. Old and recent implementation of the Starobinsky model can be found in old minimal supergravity in \cite{ceo,Ellis:2013xoa,Kallosh:2013lkr,Buchmuller:2013zfa} and in new minimal supergravity in \cite{Cecotti:1987qe, deRoo:1990zm, Ferrara:2013kca}. These theories have the common feature of adding to pure supergravity four bosonic and four fermionic degrees of freedom. At the linearized level the $4_b+4_f$ degrees of freedom comprise two massive chiral multiplets in the old minimal formulation \cite{Ferrara:1978rk} and one massive vector multiplet in the new minimal formulation. Another class of supersymmetric embedding of Starobinsky with a single chiral superfield coupled to supergravity was considered in \cite{Ketov:2010qz}. Elimination of the auxiliary field could produce $R^2$ or higher powers. However, it has been shown that \cite{Ferrara:2013wka,Ketov:2013dfa} this class of model is insufficient for realization of the Starobinsky inflation .

In this paper, we investigate the possibility of adding higher order corrections to the supersymmetric extension of Starobinsky model in the old minimal formulation. For the absence of ghosts, we consider these higher order corrections to be the supersymmetric completion of $R^n$. Utilizing $4{\rm D}$ ${\cal N}=1$ superconformal tensor calculus, we obtain the supersymmetric $R^n (n\geq 3)$ action in its component form  explicitly. Summing over all these $R^n$ actions leads to an extension of Starobinsky model to any order in terms of the power of $R$. Realization of slow roll inflation probably would restrict the higher order parameters to be sufficiently small. However, it is still interesting to explore the physical consequences implied by the higher order corrections which are accumulated during the long history of the universe.

\section{Settings}

In this section, we introduce the necessary ingredients needed in the construction of the supersymmetric $R^n$ action. The 4D $\cN=1$ superconformal tensor calculus is based on the superconformal algebra $\SU(2,2 |1)$ \cite{Ferrara:1978em}.
The gauge fields associated with general coordinate transformations $(e_\m{}^a)$, dilatations $( b_\m)$, chiral $U(1)$ symmetry $(A_\m)$ and $Q$ supersymmetry $(\p_\m^\a)$ are independent fields comprising the $12_b+12_f$ 4D $\cN=1$ superconformal Weyl multiplet. The remaining gauge fields associated with the Lorentz transformation $ \o_\m{}^{ab}$, special conformal transformation $( f_\m{}^a)$ and $S$ supersymmetry transformation $(\phi_\m^\a)$ are dependent fields. They are composite objects, which depend on the independent fields of the multiplet. We refer to \cite{freevanp} for an extended summary of the superconformal transformations of the Weyl multiplet fields, the expressions for the curvatures and other useful details. In short,
the Weyl multiplet consists of the following independent fields
\be
 \Big( e_\m{}^a,  \quad b_\m,  \quad A_\m, \quad \p_\m^{\a}\Big),
\label{GaugeFields}
\ee
while the dependent fields are given by
\bea
\o_\m{}^{ab} &=& 2 e^{\n[a} \partial_{[\m} e_{\n]}{}^{b]} - e^{\n[a} e^{b]\s} e_{\m c} \partial_\n e_s{}^c + 2 e_\m{}^{[a} b^{b]} + \ft12 \bar\p_{\m} \g^{[a} \p^{b]}  + \ft14 \bar\p^a \g_\m \p^b \,,\nn\\
\phi_\m &=& - \ft12 \g^a \widehat{R}'_{\m a}(Q) + \ft1{12} \g_\m \g^{ab} \widehat{R}'_{ab} (Q) \,,\nn\\
f_\m{}^a &=&  - \ft14 {\mathcal{R}}_\m{}^a  + \ft1{24} e_\m{}^a {\mathcal{R}} \,,
\eea
where ${\mathcal{R}}_{\m}{}^a \equiv  \widehat{R}'_{\m\n}{}^{ab}(M) e^\n{}_b$. The notation $\widehat{R}'(M)$ and $\widehat{R}'(Q)$ refers to the following expressions
\bea
\widehat{R}'_{\m\n}(Q) &=& 2 \partial_{[\m} \p_{\n]} + \ft12 \o_{[\m}{}^{ab} \g_{ab} \p_{\n]} + b_{[\m} \p_{\n]} - 3 \rmi A_{[\m} \g_5 \p_{\n]} \,,\nn\\
\widehat{R}'_{\m\n}{}^{ab}(M) &=& 2 \partial_{[\m} \o_{\n]}{}^{ab} + 2 \o_{[\m}{}^{ac} \o_{\n]c}{}^b  - \bar\p_{[\m} \g^{ab} \phi_{\n]} \,.
\eea 
 The $Q, S$ and $K$ transformation of the independent fields in the Weyl multiplet take the following form 
\bea
\d e_\m{}^a &=& \ft12 \bar\e \g^a \p_\m \nn\,,\\
\d \p_\m &=& \cD_\m  \e - \g_\m \eta  \nn\,,\\
\d b_\m &=& \ft12 \bar\e \phi_\m + 2\L_{K\m} - \ft12 \bar\eta \p_\m \nn\,,\\
\d A_\m &=& - \ft12 \rmi \bar\e \g_5\phi_\m + \ft12  \rmi \bar\eta \g_5\p_\m \,.
\label{WeylTransform}
\eea

In the superconformal method for $\cN=1$ supergravity, a compensating chiral multiplet is used to fix the extra superconformal symmetries, giving rise to the standard old minimal formulation of Poincar\'e supergravity. Chiral multiplets also provide matter coupling to conformal supergravity. Generically, a chiral multiplet consists of two complex scalars $Z$ and $\cF$, and the left-chiral projection of a Majorana spinor $P_L \chi$. Totally, these fields carry $4_b+4_f$ off shell degrees of freedom. The $Q$- and $S$- supersymmetry transformation rules of chiral multiplet are given in \cite{freevanp}
\bea
\d Z &=&  \bar\e  P_L \chi \,,\nn\\
\d P_L \chi &=&  \ft1{2}  P_L \Big( \slashed{\cD} Z + \cF \Big) \e +  \o_Z Z P_L \eta \nn\,,\\
\d \cF &=& \bar\e \slashed{\cD}  P_L \chi - 2 ( \o_Z - 1) \bar\eta P_L  \chi \,,
\label{ScalarTransformation}
\eea
where $\o_{Z}$ is the conformal weight of $Z$. The closure of supersymmetry algebra requires that the chiral $U(1)$ charge of $Z$ equals its conformal weight. The supercovariant derivatives appearing in the transformation rules takes the form \cite{freevanp}
\bea
\cD_\m Z &=& ( \partial_\m - \o_Z b_\m - i \o_Z A_\m ) Z -  \bar\p_\m P_L  \chi  \ \,,\nn\\
\cD_\m  P_L \chi &=& P_L \Big[ \Big(\partial_\m - (\o_Z + \ft12) b_\m + \ft14 \o_\m{}^{ab} \g_{ab} \nn\\
&& - (\o_Z - \ft32 ) \rmi A_\m \Big) \chi   - \ft12 \Big( \slashed{\cD} Z + \cF \Big) \p_\m - \o_Z Z  \phi_\m \Big] \,.
\label{ConformalDerivative}
\eea
Since a chiral multiplet is identified by its first component, we therefore use its first component to stand for the whole multiplet.

\section{Construction of the Supersymmetric $R^n$ Action}
To construct the supersymmetric completion of $R^n$, we first introduce the component form of fusion rule satisfied by the chiral multiplets. Based on the fusion rule, one can construct a composite chiral multiplet by using two known chiral multiplets. Given the fact that chiral multiplets in 4D ${\cal N}=1$ conformal supergravity can have arbitrary Weyl weights, fusion rule is able to be utilized flexibly which facilitates the supersymmetrization of $R^n$ tremendously. Without referring to the cumbersome superspace formulation, we obtain the bosonic part of supersymmetric $R^n$ action straightforwardly. In particular, we attain the supersymmetric $R^3$ action which is beyond the construction adopted in \cite{Ferrara:2013kca,Farakos:2013cqa}.

In order to achieve the $R^n$ supergravity, we utilize four different chiral multiplets, which are given in Table \ref{T1} along with the Weyl weights and chiral $U(1)$ charges of their lowest components.
\begin{table}[h]
\centering
\caption{\label{special}\footnotesize  List of Chiral multiplets used in the construction of the $R^n$ extension of the Starobinsky Model.}
\begin{tabular}{|c|c|c|c|}
\hline
\textbf{Name}& Weyl weight& Chiral weight&Components\\
\hline
Neutral& $0$& 0&$(\s, P_L\p, F)$\\
\hline
Compensating& $1$&1&$(\phi, P_L\l, S)$\\
\hline
Curvature& $2$&2&$(\bar S,\slashed{\cD} P_R \l, \Box^C\bar{\phi})$\\
\hline
Auxiliary& $3$&3&$(A, P_L\O, N)$\\
\hline
\end{tabular}
\label{T1}
\end{table}
From (\ref{ConformalDerivative}), we can see that their supersymmetry transformation rules are determined by the Weyl weights and chiral $U(1)$ charges of the lowest components. The highest weight component of the auxiliary multiplet has Weyl weight 4, therefore its superconformal completion can be used as the Lagrangian density. Its full expression is given in \cite{freevanp}
\be
e^{-1} \cL_{A} = \text{Re} \Big( N +  \bar\p_\m \g^\m P_L \O + \ft12  A \bar\p_\m \g^{\m\n} P_R  \p_\n \Big) \,.
\label{Weight4Action}
\ee
\subsection{Fusion Rule and the Map Between a Chiral and the Weyl Multiplet}
\begin{figure}[H]
\centering
\caption{Fusion of two chiral multiplets}
\tikzstyle{place}=[circle,thick]
\tikzstyle{transition}=[rectangle,draw=black!50,fill=black!20,thick]
\begin{tikzpicture}[inner sep=1.5mm]
\node[circle,fill=blue!20] at (0.5,2) (Z2) [place] {$Z_2$};
\node[circle,fill=yellow!20] at (-1,1) (Z1) [place] {$Z_1$};
\node[circle,fill=green!20] at (2,1) (Z3) [place] {$Z_3$};
\draw [-,thick] (Z2) -- (0.5,1);
\draw [-,thick] (Z1) -- (0,1);
\draw [->,thick] (0,1) -- (Z3);
\label{D1}
\end{tikzpicture}
\end{figure}
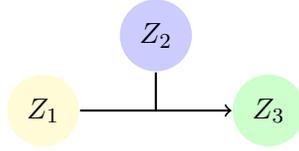

In superspace formulation, the multiplication of two chiral superfield gives rise to a chiral superfield. In terms of components, this means \cite{freevanp}
\bea
Z_3 &=& Z_1Z_2 \,,\nn\\
P_L\chi_3 &=& Z_1P_L\chi_2 + Z_2P_L\chi_1 \,,\nn\\
\cF_3 &=& Z_1\cF_2 + Z_2\cF_1  - 2 \bar\chi_1 P_L \chi_2 \,,
\label{L1L2Map1}
\eea
where the conformal weight of $Z_3$ equals the sum of the conformal weights of $Z_1$ and $Z_2$. This fusion rule is visualized in Fig(\ref{D1}).
It can be deduced from (\ref{L1L2Map1}) that when multiple chiral multiplets are composed into a new chiral multiplet, the order of the fusion will not affect the result. In other words, if $Z_{i_1\cdots i_n}$ results from the successive fusions of $Z_{i_1}$, $Z_{i_2}\cdots Z_{i_n}$, then $Z_{i_1\cdots i_n}$ is totally symmetric with respect to indices $i_1\cdots i_n$.

\begin{figure}[H]
\centering
\caption{The formaton of neutral multiplet based on aa curvature multiplet $(\bar S',\slashed{\cD} P_R \l', \Box^C\bar{\phi'})$ and 2 compensating multiplets.}
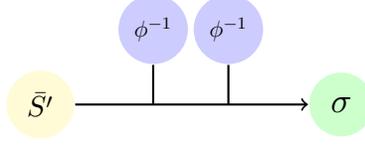
\tikzstyle{place}=[circle,thick]
\tikzstyle{transition}=[rectangle,draw=black!50,fill=black!20,thick]
\begin{tikzpicture}[inner sep=1.5mm]
\node[circle,fill=blue!20,scale=0.8] at (0,2) (z1) [place] {$\phi^{-1}$};
\node[circle,fill=blue!20,scale=0.8] at (1,2) (z4) [place] {$\phi^{-1}$};
\node[circle,fill=yellow!20,scale=1] at (-1.5,1) (z2) [place] {$\bar{S'}$};
\node[circle,fill=green!20,scale=1.2] at (2.5,1) (z3) [place] {$\sigma$};
\draw [-,thick] (z1) -- (0,1);
\draw [-,thick] (z2) -- (0,1);
\draw [->,thick] (0,1) -- (z3);
\draw [-,thick] (z4) -- (1,1);
\label{D2}
\end{tikzpicture}
\end{figure}

The fusion rule (\ref{L1L2Map1}) can be inverted and lead to another fusion rule for chiral multiplet
\bea
Z_1 &=& Z_3Z_2^{-1} \,,\nn\\
P_L\chi_1 &=& Z_2^{-1}P_L\chi_3 - Z_3Z_2^{-2}P_L\chi_2 \,,\nn\\
\cF_1 &=& Z_2^{-1}\cF_3 - Z_3Z_2^{-2}\cF_2  + 2Z_2^{-2}\bar\chi_2 P_L \chi_3\nn\\
&&-2Z_3Z_2^{-3}\bar\chi_2 P_L \chi_2 \,.
\label{L1L2Map2}
\eea
By using the second fusion rule (\ref{L1L2Map2}), a neutral multiplet $\s$ can be formed with the aid of 2 compensating multiplets $\phi$ and a curvature multiplet $(\bar S',\slashed{\cD} P_R \l', \Box^C\bar{\phi'})$ based on a weight-1 chiral multiplet with field contents ($\phi'$, $P_L \l'$, $S'$). 
Here the ``primed'' multiplet can be generically different from the compensating multiplet denoted by ($\phi$, $P_L \l$, $S$) which will be utilized to fix the extra superconformal symmetries. The formation of the composite neutral multiplet is illustrated in Fig(\ref{D2}).
 As a consequence, the expressions for the composite fields are
\bea
\s&=&\phi^{-2} \bar S',\quad  P_L\p=\phi^{-2}\slashed{\cD} P_R \l'-2\bar S'\phi^{-3}P_L\l,\nn\\
F&=&\phi ^{-2}\Box^C\bar{\phi'} - 2\phi^{-3}\bar S' S+2\bar S'\phi^{-4}\bar\l P_L\l\nn\\
\qquad&&+4\phi^{-3}\bar\l P_L\slashed{\cD} \l'-8\bar S'\phi^{-4}\bar\l P_L\l.
\label{01Embed1}
\eea

If the anti-chiral multiplet $\bar\phi'$ happens to be the complex conjugate of the compensating multiplet $\phi$, we can derive a map between the neutral multiplet and the Weyl multiplet after eliminating the extra superconformal symmetry by gauge fixing.

We choose the following gauge conditions to fix dilatation symmetry, local chiral $U(1)$ symmetry, $S$ supersymmetry and special conformal symmetry,
\be
\phi =\sqrt{3},\quad \l =0, \quad b_{\mu}=0,
\label{GaugeFixing}
\ee
which also leads to the map from the neutral multiplet to the supergravity multiplet. After an overal sign change of the fields in the composite neutral multiplet, namely, $\sigma\rightarrow-\sqrt{3}\sigma$, $P_L\psi\rightarrow-\sqrt{3}P_L\psi$ and $F\rightarrow-\sqrt{3}F$, we obtain
\bea
\s &=& - \ft1{\sqrt3} \bar{S},\quad  P_L\p=-\ft16P_L\g^{\m\n}\widehat{\p}_{\m\n} \,,\nn\\
F &=& \ft16 {\mathcal{R}}+ A_\m A^\m  - \rmi\cD_a A^a+ \ft23 \bar{S} S + \ft1{\sqrt3} \bar\p^aP_R\cD_a\l\,,
\label{MainMap}
\eea
where ${\p}_{\m\n}$ is the covariant curvature of gravitino defined by
\bea
\widehat{\p}_{\m\n} &=&  2 \Big( \partial_{[\m} + \ft12 b_{[\m} + \ft14 \o_{[\m}{}^{ab} \g_{ab} - \ft32 \rmi A_{[\m} \g_5 \p_{\n]} - \ft12 \rmi \g_{[\m} \slashed{A}\g_5 \p_{\n]} \nn\\
\qquad &&+\ft1{2\sqrt3} \g_{[\m}(\rm{Re}S+\rmi \rm{Im}S\g_5)   \p_{\n]}  \Big)\,. 
\eea
$R$ enters $F$ through $\Box^C \bar{\phi}$ and the resulting chiral multiplet is only manifestly Poincar\'e invariant.
It should be noted that $F$ contains the Ricci scalar, and $S$ now becomes the auxiliary scalar in the old minimal Poincar\'e supergravity, which is stated by its transformation rule
\be
\d S = \ft{1}{2\sqrt3} \bar\e P_R \g^{\m\n} \widehat{\p}_{\m\n} \,.
\ee
The 4D old minimal supergravity multiplet comprises the fields $( e_\m{}^a, \p_\m^a,  S,  A_\m)$ which transfom under supersymmetry as
\bea
\d e_\m{}^a &=& \ft12 \bar\e \g^a \p_\m \,,\nn\\
\d  \p_\m &=&\Big( \partial_\m + \ft14 \o_\m{}^{ab} \g_{ab} - \ft32 \rmi A_\m \g_5 \Big) \e - \ft12 \rmi \g_\m \slashed{A}\g_5 \e + \ft1{2\sqrt3} \g_\m(\rm{Re}S+\rmi \rm{Im}S\g_5)  \e \nn\,,\\
\d A_\m &=& - \ft14 \rmi \bar\e \g_5 \Big( - \g^\n \widehat{\p}_{\m\n} + \ft16 \g_\m \g^{ab} \widehat{\p}_{ab} \Big)  \,,\nn\\
\d S &=& \ft{1}{2\sqrt3} \bar\e P_R \g^{\m\n} \widehat{\p}_{\m\n} \,.
\label{OffShellTransformationRules}
\eea

The supersymmetry transformation rule of 4D old minimal supergravity (\ref{OffShellTransformationRules}) can be deduce from the transformation rule of 4D $\cN=1$ conformal supergravity (\ref{WeylTransform}) as a consequence of gauge fixing conditions (\ref{GaugeFixing}). One should notice that in order to maintain the gauge fixing condition (\ref{GaugeFixing}), the compensating transformations are needed and their transformation parameters are determined as follows
\bea
\L_D &=& \L_{U(1)} = 0 \,,\nn\\
\L_{K\m} &=& - \ft14 \bar\e \phi_\m + \ft14 \bar\eta \p_\m \,,\nn\\
P_L \eta &=& \ft12 \rmi P_L \slashed{A}  \e - \ft1{2\sqrt3} S P_L \e  \,.
\label{DecompostionRules}
\eea
Since the transformation rule (\ref{OffShellTransformationRules}) is off-shell, its closure property does not rely on the detailed form of the Lagrangian. The commutator of the supersymmetry transformation rule  (\ref{OffShellTransformationRules}) can be found in \cite{Ferrara:1978em}.

\subsection{Supersymmetric Completion of $R^n$ Action}
With previous preparations, we now proceed with the supersymmtrization of $R^n$ action.
We will utilize the fact that the neutral chiral multiplet $\s$ carries 0 Weyl weight and chiral charge. Therefore, according to the fusion rule (\ref{L1L2Map1}), it can be combined with a second chiral multiplet to form a third chiral multiplet which shares the same properties with the second chiral multiplet. This fusion can be duplicated for arbitrary times. At the end of each fusion, we arrive at a different composite multiplet, however, the conformal weight and chiral $U(1)$ charge remain unchanged. Considering that the superconformal action (\ref{Weight4Action}) is associated with the auxiliary multiplet, we need to construct the auxiliary multiplet in terms of the neutral multiplet and the compensating multiplet. The former brings the $R$ dependence through $F$, while the latter balances the conformal weight and chiral charge. By a process of trial and error, we find that to obtain the supersymmtrization of $R^n$, two independent neutral multiplets are needed. Fusing 3 compensating multiplets $\phi$, $n-1$ of the first neutral multiplets denoted by $\s'$ and 1 of the second neutral multiplet represented by $\s$ leads to the composite expression for the auxiliary multiplet
\begin{figure}  [H]
\centering
\caption{Fusion rule of an auxiliary multiplet for the construction of $R^n$ action}
\tikzstyle{place}=[circle,thick]
\tikzstyle{transition}=[rectangle,draw=black!50,fill=black!20,thick]
\begin{tikzpicture}[inner sep=1.5mm]
\node[circle,fill=blue!20] at (0,2) (Z1) [place] {$\phi$};
\node[circle,fill=blue!20] at (0,0) (Z2) [place] {$\phi$};
\node[circle,fill=blue!20] at (-1,1) (Z4) [place] {$\phi$};
\node[circle,fill=red!20,scale=1.15] at (1,2) (M1) [place] {$\s$};
\node[circle,fill=green!20] at (2,0) (N1) [place] {$\s'$};
\node[circle,fill=green!20] at (4,0) (N2) [place] {$\s'$};
\node[circle,fill=gray!20] at (5,1) (A1) [place] {$A$};
\draw [-] (Z1) -- (Z2);
\draw [-] (Z4) -- (0,1);
\draw [-] (0,1) -- (1.0,1);
\draw [-] (M1) -- (1.0,1);
\draw [-] (N1) -- (2.0,1);
\draw [-] (N2) -- (4,1);
\draw [-] (1,1)-- (2,1);
\draw [-,dashed]  (2,1)--(4,1);
\draw [->,thick] (4,1)-- (A1);
\draw [<-,thick] (2,0.7)-- (2.7,0.7);
\node at (3.0,0.7) {$n$};
\draw [->,thick] (3.3,0.7)-- (4,0.7);
\label{D3}
\end{tikzpicture}
\end{figure}
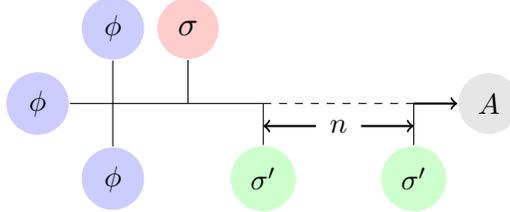
\bea
 A& =&\phi^3 \s (\s')^{n - 1} ,\quad P_L\Omega=(n-1)\s\phi^3(\s')^{n-2}P_L\psi'+(\s')^{n-1}P_L(3\s\phi^2\l+\phi^3\psi),\nn\\
 N &=& (\s')^{n-1} ( 3 \phi^2 \s S + \phi^3 F)+ (n-1) (\s')^{n-2} \s \phi^3 F'-2(n-1)\s(\s')^{n-2}\phi^3\bar\psi'P_L\psi'\nn\\
\qquad&&-2(n-1)(\s')^{n-2}\phi^3\bar\psi P_L\psi'-6(n-1)\s(\s')^{n-2}\phi^2\bar\l P_L\psi'\nn\\
\qquad &&-6\s(\s')^{n-1}\phi\bar\l P_L\l -6(\s')^{n-1}\phi^2\bar\l P_L\psi.
\eea
Then, using the auxiliary chiral multiplet action (\ref{Weight4Action}), we obtain the corresponding Lagrangian as
\bea
e^{-1} \cL &=&\text{Re}\Bigg[ (\s')^{n-1}\Big( 3 \phi^2 \s S + \phi^3 F+3\s\phi^2\bar\psi^{\m}\g_{\m}P_L\l+\phi^3\bar\psi^{\m}\g_{\m}P_L\psi-6\s\phi\bar\l P_L\l\nn\\
&&\qquad-6\phi^2\bar\l P_L\psi+\ft12\s\phi^3\bar\psi^{\mu}\g^{\m\n}P_R\psi_{\n} \Big)+(n-1)(\s')^{n-2}\Big(\s\phi^3 F'-2\phi^3\bar\psi P_L\psi'\nn\\
&&\qquad-6\s\phi^2\bar\l P_L\psi'+\s\phi^3\bar\psi^{\m}\g_{\m}P_L\psi'-2\s\phi^3\bar\psi' P_L\psi'\Big)\Bigg].
\label{FSCA}
\eea
In the above expressions, $F$ can be replaced according to (\ref{MainMap}) which includes the Ricci scalar. Thus, if the neutral multiplet $\s'$ can be realized in terms of the neutral multiplet $\s$ in such a way that $\s'$ depends on $F$ linearly, we are capable of deriving the supersymmetric $R^n$ action. This can be achieved by two steps.
We first combine the neutral multiplet $\s$ with the compensating multiplet $\phi$ to form a new weight-1 chiral multiplet $\phi'$
\be
\phi'= \sigma \phi, \quad P_L\l'=\s P_L\l+\phi P_L\psi,\quad  S' = \s S + \phi F-2\bar\l P_L\psi.
\label{weight1}
\ee
We then substitute (\ref{weight1}) into $(\ref{01Embed1})$.
Finally, the neutral multiplet $\s'$ is given as the composite built upon the neutral multiplet $\s$ and the compensating multiplet $\phi$
\bea
\s' &=& \phi^{-2} (\bar\s \bar{S} + \bar\phi \bar{F} )+ ({\rm \l - terms}) \,,\nn\\
P_L\psi' &=& \phi^{-2}\slashed{\cD}P_R(\s\l+\phi\psi)+  ({\rm \l - terms}) \nn\\
F' &=& \phi^{-2} \Box^C (\bar\phi \bar\s) - 2 \phi^{-3} \bar\s S \bar{S} - 2 \phi^{-3} \bar\phi S \bar{F}+  ({\rm \l - terms}) \,,
\label{FC}
\eea
where $\s'$ depends on $F$ linearly as required, and the terms explicitly depending on $\l$ are ommited since they vanish upon the gauge fixing (\ref{GaugeFixing}). Plugging (\ref{FC}) into (\ref{FSCA}), fixing the redundant superconformal symmetry and replacing $\s$, $F$ according to (\ref{MainMap}) completes the supersymmetrization of $R^n$ action. 
For $n=1$, (\ref{Rnaction}) gives rises the old minimal Poincar\'e supergravity
\be
e^{-1} \cL_{R}=R(\omega(e,\psi))-\bar\psi_{\m}\g^{\m\n\r}(\partial_{\n}+\ft14\omega_{\n}^{~ab}(e,\psi)\g_{ab})\psi_{\r}+6 A^\m A_\m-2S\bar{S}.
\label{Raction}
\ee
For $n=2$, (\ref{Rnaction}) reproduces the supersymmetric $R^2$ action in old minimal supergravity \cite{ceo}
\be
e^{-1} \cL_{R^2}
= R^2 + 12 R  A^\m A_\m + 2 R \bar{S} S + 36 (\nabla_\m A^\m)^2 - 12 D_\m \bar{S} D^\m S+4 (S\bar{S} + 3  A^\m A_\m)^2  \,,
\label{R2action}
\ee
where we have defined $D_\m S\equiv\partial_\m S+{\rm i}A_\m S$. To arrive at (\ref{Raction}) and (\ref{R2action}), an overall scaling of the Lagrangian has been performed.
For general $n$, the purely bosonic part ofsupersymmetric $R^n$ action takes the following form
\footnote{Having posted this paper, we were informed by S.Ferrara that a superspace formulation of the  $R^n$ action can be found in \cite{ceo}. However, in \cite{ceo}, the component form of the $R^n$ action was not computed explicitly. In this paper, we derive the supersymmetric $R^n$ action by superconformal tensor calculus and present the supersymmetric $R^n$ action explicitly. Also, the lower power of $R$ terms in the  $R^n$ action can be affected by adding the $R^p(S^{q-1}+\bar{S}^{q-1})$ type invariants given in eq.(\ref{RSaction}).}
\bea
e^{-1} \cL_{R^n} &=& \text{Re} \Bigg[ \Big(R + 6 A^\m A_\m + 6 \rmi \nabla_\m A^\m + 2 S \bar{S} \Big)^{n-2} \times\Big( \ft1{12} R^2 + R A^\m A_\m+ \ft{n - 1}{6} R S \bar{S}\nn\\
&&\quad + 3 (A^\m A_\m)^2 + \ft{2n - 3}3(S\bar{S})^2 + ( n - 1) A^\m A_\m S \bar{S}+ 3 (\nabla_\m A^\m)^2+(n-1) \bar{S} \Box S   \nn\\
&& \quad + (3n - 5) \rmi S \bar{S} (\nabla_\m A^\m) + (2n - 2) \rmi A_\m \bar{S} \partial^\m S \Big) \Bigg] \,. 
\label{Rnaction}
\eea
The full Lagrangian including the gravitino contribution can be obtained straightforwardly by inserting the expressions (\ref{MainMap}) and (\ref{FC}) into (\ref{FSCA}). From the above results, we can see that around the supersymmetric Minkowski vacuum, the $R^n$ action with $n\ge3$ does not modify the spectrum of the theory but only the interactions. Therefore, the spectrum around the Minkowski vacuum can be studied by simply focusing on the supersymmetric $R+R^2/(6M^2)$ action. It turns out that besides the massless supergravity multiplet, there are two massive chiral multiplets. The first  massive chiral multiplet consists of the complex scalar $S$, and two components of the $\g^{\mu}\psi_{\mu}$, the second massive multiplet is comprised of $\nabla^{\m}A_{\mu}$, the spin-0 part of metric and the remaining two components of $\partial^{\mu}\psi_{\mu}$. Different from the ordinary two derivative theory, $\partial^{\mu}\psi_{\mu}$ becomes dynamical in higher-derivative theories as can be seen  in supersymmetric $R+R^2/(6M^2)$ model, Rarita-Schwinger equation takes the following form
\be
-\g^{\m\n\r}\partial_{\n}\psi_{\r}+\frac{2}{3M^2}\g^{\m\n}\partial_{\n}\slashed{\partial}\g^{\r\l}\partial_{\r}\psi_{\l}=0.
\ee
Adopting the gauge choice $\g^{\mu}\psi_{\mu}=0$, we obatin 
\be
(\Box-M^2)\partial^{\m}\psi_{\m}=0,
\ee
which reveals the fact that the four components of $\partial^{\m}\psi_{\m}$ are dynamical degrees of freedom with mass squared $M^2$.
\subsection{Cosmological Constant}
The supersymmetric cosmological constant term can be derived by constructing an auxiliary multiplet using 3 compensating multiplets. This amounts to set $\s=\s'=1$ and $F=F'=0$ in (\ref{FSCA}). Upon eliminating the extra conformal symmetries, the supersymmetric cosmological constant term takes the simple form
\be
e^{-1}{\cal L}_{\Lambda}= \text{Re}(S).
\ee
The vacuum expectation value of $S$ behaves as a cosmological constant parameter appearing in the Lagrangian. However, in a higher curvature gravity theory, the higher derivative interactions can generate cosmological constant. Thus the effective cosmological constant should be read off from the value of the curvature tensor solved from the equation of motion.

\section{Supersymmetric Vacua and $R^n$ Extended Minimal  Starobinsky Model}

In this section, we investigate the supersymmetric vacua, and present the minimal Volkov-Akulov formulation of $R^n$ extended Starobinsky model. 

\subsection{Supersymmetric Vacua in Polynomial $f(R)$ Supergravity}
In this section, we consider the supersymmetric vacuum solution in polynomial $f(R)$ supergravity with the following Lagrangian
\be
{\cal L}=\xi_0{\cal L}_{\Lambda}+\sum_{n\geq1}\xi_n {\cal L}_{R^n}.
\label{}
\ee
The supersymmetric vacuum solutions are determined by the vanishing of the gravitino variation
\be
\delta\psi_{\mu}\equiv (\partial_{\mu}+\ft14\omega_{\mu}^{ab}\gamma_{ab}-\ft32{\rm i}A_{\mu}\gamma_5)\epsilon-\ft{\rm i}{2}\gamma_{\mu}\slashed{A}\gamma_5\epsilon+\ft1{2\sqrt{3}}\gamma_{\mu}({\rm Re}S+{\rm i Im}S\gamma_5)\epsilon=0.
\ee
We make the following ansatz for the vacuum solutions
\be
R_{\mu\nu\r\s}=\frac{a}{12} (g_{\mu\rho}g_{\nu\sigma}-g_{\mu\sigma}g_{\nu\rho}),\quad S=b,\quad A_{\mu}=0,
\ee
with $a$ and $b$ being two real constants. The integrability condition from the vanishing of gravitino transformation requires 
\be
a=-b^2.
\ee
Using this equation, we find that the full set of equations of motion can be reduced to a polynomial equation satisfied by $b$
\be
\xi_0+\sum_{n\geq1}(-2)^{n-1}\frac{n-2}3\xi_nb^{2n-1}=0.
\ee
From the above formula, we can see that to have supersymmetric Minkowki vacuum, $\xi_0$ has to be 0.
\subsection{ The $R^n$ Extended Minimal Volkov-Akulov-Starobinsky Model}
The minimal Volkov-Akulov-Starobinsky model was constructed recently in
\cite{Antoniadis:2014oya} by using a non-linear realizationof neutral chiral multiplet. If we denote the superfield corresponding to this multiplet by $X$,
then $X$ satisfies the constraint automatically \cite{Rocek:1978nb}-\cite{Komargodski:2009rz}
\be
X^2=0.
\ee
In terms of the components, this means the lowest component in the neutral multiplet becomes bilinear in $P_L\chi$. In the Weyl representation, $P_L\chi$ is a two component spinor, and the constrained neutral multiplet can be expressed as
\be
X=\frac{(P_L\chi)^{\a}(P_L\chi)_{\a}}{F}+2\theta P_L\chi+\theta^2 F.
\ee
Inserting the neutral muliplet given in (\ref{MainMap}), we obtain a constraint neutral multiplet based on the fields in the old minimal supergravity multiplet as follows
\bea
&&\s=-\frac{\bar{S}}{\sqrt{3}}=\frac{(P_L\chi)^{\a}(P_L\chi)_{\a}}{F}\,,\nn\\
&&P_L\chi=-\ft16 P_L\gamma^{ab}\widehat{\psi}_{ab},\nn\\
&&F=\ft16 R + A_\m A^\m  - \rmi (\nabla_\m A^\m)+{\rm fermions} ,
\eea
where $\widehat{\psi}_{ab}$ is the field strength of gravitino. The above constrained neutral multiplet implies that the scalar auxiliary field in old minimal supergravity can be realized as the bilinear of gravitino field strength. If using this multiplet to construct the supersymmetric $R^n$ action, we can readily write down the bosonic part of the minimal supergymmetric $R^n$ action by simply omitting terms depending on $S$, since $S$ is bilinear in fermions
\be
e^{-1}{\cal L}^{\rm min}_{R^n}=\text{Re} \Bigg[ \Big(R + 6 A^\m A_\m + 6 \rmi \nabla_\m A^\m \Big)^{n-2} \Bigg] \times\Big( \ft1{12} R^2 + R A^\m A_\m+ 3 (A^\m A_\m)^2+ 3 (\nabla_\m A^\m)^2\Big)\label{mRn} .
\ee
The minimal supergymmetric $R^n$ theory only has Minkowski metric as the supersymmetric vacuum solution because the scalar field $S$ must vanish as a consequence of the vanishing gravitino in the vacuum.  
\section{The Dual Model}
The $R^n$ extended Starobinsky model possesses a dual description in terms of the standard supergravity coupled to two chiral multiplets model. The dual model corresponding to (\ref{RnStar}) can be derived as follows. After plugging (\ref{FC}) into (\ref{FSCA}) and fixing the redundant superconformal symmetries, we do not replace $\s$, $F$ according to (\ref{MainMap}) undoing the step leading to the $R^n$ action. Then Lagrangian multipliers should be added to the Lagrangian whose equations of motion imply the map from chiral multiplet $\s$ to the supergravity multiplet (\ref{MainMap}). Integrating out the auxiliary fields of the supergravity multiplet and performing a Weyl scaling on the metric eventually gives rises to the standard supergravity coupled to two chiral multiplets. 

Generically, the Lagrangian of the dual model describing supergravity coupled to two chiral multiplets is very complicated and it is hardly to discuss the physical consequence of this model. However, we notice that the supersymmetric $R^n$ Lagrangian eq.(\ref{Rnaction}) is invariant under $S\rightarrow-S$. Therefore, we can consistently set $S=0$. The resulting Lagrangian takes the same form as the minimal Volkov-Akulov $R^n$  Lagrangian eq.(\ref{mRn}). This truncated $R+R^2+R^n$  model can then be dualized to a scalar coupled supergravity model as follows
\bea
\kappa^2e^{-1}{\cal L}_{R+R^2+R^n}&=&{\rm Re}\Big(2F+\frac{2}{3M^2}F\bar{F}+\frac{2^{n-1}\lambda_n}{n(3M^2)^{n-1}}F\bar{F}^{n-1}+\ft12 T R \nn\\
&&\qquad +3T A^{\mu}A_{\mu}-3{\rm i}T\nabla^{\mu}A_{\mu}-T F \Big)\,.\label{action1}
\eea
It can be seen that the $R^n$ extended Starobinsky model can be recovered by integrating out $T$. If instead, $F$ and $A_{\mu}$ is integrated out, the above Lagrangian gives rise to the scalar-tensor theory.  We introduce the following notation for the complex scalars $T$  and $F$  by
 \be
 T=e^{\kappa\phi\sqrt{\frac{2}{3}}}+i\chi\sqrt{\frac{2}{3}},\quad F=F_R+{\rm i} F_I.
 \ee
Integrating out $F$ and $A_{\mu}$ followed by a Weyl scaling on the metric 
\be
g_{\mu\nu}\rightarrow e^{-\kappa\sqrt{\frac{2}{3}}\phi}g_{\mu\nu},
\ee
leads to the dual model 
\be
e^{-1}{\cal L}=\frac1{2\kappa^2}R-\ft12(\partial\phi)^2-\ft12e^{-2\kappa\phi\sqrt{\frac{2}{3}}}(\partial\chi)^2-V(\chi,\phi),\label{dualRn}
\ee
where the scalar potential is given by
\bea
V(\chi,\phi)&=&e^{-2\sqrt{\frac{2}{3}}\kappa\phi}{\rm Re}[2F+\frac{2}{3M^2}F\bar{F}+\frac{2^{n-1}\lambda_n}{n(3M^2)^{n-1}}F\bar{F}^{n-1}],\nn\\
e^{\kappa\phi\sqrt{\frac{2}{3}}}&=&2+\frac4{3M^2}F_R+\frac{2^{n-1}\lambda_n}{n(3M^2)^{n-1}}\Big(2F_R{\rm Re}\bar{F}^{n-2}+(F_R^2+F_I^2)\partial_{F_R}{\rm Re}\bar{F}^{n-2}\Big),\nn\\
-\chi\sqrt{\frac{2}{3}}&=&\frac4{3M^2}F_I+\frac{2^{n-1}\lambda_n}{n(3M^2)^{n-1}}\Big(2F_I{\rm Re}\bar{F}^{n-2}+(F_R^2+F_I^2)\partial_{F_I}{\rm Re}\bar{F}^{n-2}\Big).
\eea
Since ${\rm Re}\bar{F}^{n-2}$ contains only the even power of $F_I$, we can consistently choose $F_I=0$ and $\chi=0$. In this case, the model is equivalent to the non-supersymmetric $R^n$ extended Starobinsky model studied in \cite{Huang:2013hsb}. According to \cite{Huang:2013hsb}, slow-roll inflation can be achieved in the region where $\phi$ satisfies
\be
\lambda_n(e^{\frac{2}{3}\kappa\phi}-1)^{n-2}\ll1.
\ee
The spectral index and tensor-to-scalar ration are respectively given by
\be
n_s\simeq1-\frac2N(1+\delta),\quad r\simeq\frac{12}{N^2}\Big[1-\frac2{n-1}\delta\Big],
\ee
with
\be
\delta=\frac{(n-1)^2(n-2)}{2n^2}\Big(\frac43N\Big)^{n-1}.
\ee
PLANCK \cite{Ade:2013uln} and WMAP \cite{Komatsu:2014ioa} predict
\be
n_s= 0.960\pm  0.007.
\ee
This means for e-folding number $N=50$, the contribution from the $R^n$ term must be very small $\delta\ll1$. Therefore the 
scalar-to-tensor ratio in the $R^n$ extended Starobinsky model is still around 0.005. This value is not large enough to explain the BICPE2 result \cite{Ade:2014xna} suggesting
\be
r=0.16^{+0.06}_{-0.05}.
\ee
 In the two scalar model eq.(\ref{dualRn}), some other possible inflation trajectory has been considered which leads to, for instance, the imaginary Starobinksy model \cite{Ferrara:2014ima}.  In the Imaginary Starobinsky model, $\chi$ is considered as the inflaton and the initial value of $\phi$ is chosen to be 0. If $\phi$ is stabilized around 0, \cite{Ferrara:2014ima} shows that large vaule of $r$ can be obtained along the inflation trajectory driven by $\chi$.  However, as pointed out by \cite{Kallosh:2014qta}, $\phi$ is not stabilized around 0 and in fact, it runs away from 0 to a significantly large value during a rather short period. As a consequence, the effective potential of $\chi$ is deformed and the dominant inflation trajectory is back to the one driven by $\phi$. As discussed before, the $\phi$-inflation trajectory tends to give small value of $r$.
\section{Discussion}

We have constructed the supersymmetric completion of $R^n$ action in the old minimal formulation of 4D $\cN=1$ supergravity, extending the Staronbinsky model to arbitrary power of $R$ in a supersymmetric pattern. The action is given by
\be
e^{-1}{\cal L}=R+e^{-1}{\cal L}_{R^2}/(6M^2)+\sum_{n\geq3}\xi_n e^{-1}{\cal L}_{R^n},
\label{RnStar}
\ee
where the supersymmetric $R^n$ action can be found in (\ref{Rnaction}). The construction used 3 compensating multiplets $\phi$, $n-1$ of neutral multiplet $\s'$ and 1 neutral multiplet $\s$. Instead, if considering $p-1$ number of $\s'$ and $q$ number of $\s$, one can obtain the dilatonic $R^n$ action as 
\be
e^{-1} \cL_{R^pS^{q-1}} =\text{Re} \Bigg[ (\s')^{p-1} ( 3 \phi^2 \s^q S + q\phi^3\s^{q-1} F)+ (p-1) (\s')^{p-2} \s^{q} \phi^3 F' \Bigg],
\label{RSaction}
\ee
with $\s',F'$ being replaced by (\ref{FC}) and $\s, F$ being further replaced by (\ref{MainMap}). It then can be seen that in (\ref{RSaction}) the term with highest power of $R$ is given by $R^p(S^{q-1}+\bar{S}^{q-1})$. Adding these dilatonic $R^n$ invariants to the action (\ref{RnStar}) may bring novel feature to the inflation model which is worth a future study.

In the $R^n$ extended Starobinsky model, the scalars participating inflation are the spin-0 part of the metric, the longitudinal mode of $A_{\m}$ and the complex scalar $S$. The Lagrangian (\ref{Rnaction}) seems to suggest that rather complicated interactions are involved among these scalars. In terms of the dual model, this would imply very sophisticated form of the scalar potential. The polynomial $f(R)$ extended model (\ref{RnStar}) contains infinite number of parameters, thus its dual model is capable of describing nearly arbitrary inflation potentials. The complex scalar $S$ and the longitudinal mode of $A_{\m}$ can be stabilized around 0 and the resulting theory is equivalent to the non-supersymmetric $R^n$ extended Starobinsky model studied in \cite{Huang:2013hsb} . Generic inflation trajectory with $S$ being turned on could be interesting and deserves a future study.

In this paper we only focus on the model with a $R^n$
correction term to the Starobinsky's inflation model. Actually
some other correction terms, such as $R_{\mu\nu}R^{\mu\nu}$,
$R_{\mu\nu\rho\lambda}R^{\mu\nu\rho\lambda}$ and derivatives of the curvatures, are also expected in the UV complete
quantum theory of gravity. However, generically, the inclusion of any finite number of the other higher derivative terms will essentially introduce ghosts and cause instability.  An infinite sum of the higher derivative terms might result in a covariant and ghost free theory of gravity as investigated in \cite{Biswas:2011ar, Biswas:2013dry, Chialva:2014rla}. Since such model contains an infinite series of higher derivative terms,  it would be interesting and challenging to implement supersymmetry.

\section*{Acknowledgements}
We are grateful to Sergio Ferrara for useful communications. The work of M.O is funded by a grant from Groningen University and the KNAW. The work of Y.P. is supported in part by DOE grant DE-SC0010813.

\end{document}